\documentclass{acmsmall}
\let\subcaption\relax

\usepackage[ruled]{algorithm2e}
\renewcommand{\algorithmcfname}{ALGORITHM}
\SetAlFnt{\small}
\SetAlCapFnt{\small}
\SetAlCapNameFnt{\small}
\SetAlCapHSkip{0pt}
\IncMargin{-\parindent}

\usepackage{graphicx}
\usepackage{subcaption}
\usepackage{stfloats}
\usepackage{listings}
\usepackage{hyperref}
\usepackage{multirow}

\begin{document}

\markboth{G. Inggs}{Algorithmic Trading: A brief, computational finance case study on data centre FPGAs}

\title{Algorithmic Trading: \\ A brief, computational finance case study on data centre FPGAs}
\author{GORDON INGGS
\affil{Circuit and Systems Research Group, Department of Electrical and Electronic Engineering, Imperial College London}
}

\begin{abstract}
Increasingly FPGAs will be deployed at scale as a result of the need for increased power efficient computation and improved high level synthesis tool flows, creating a new category of device: data centre FPGAs. A method for using these FPGAs is to identify what proportion of a given workload would benefit from being implemented upon the available FPGAs while minimising communication off-chip. As part of the implementation of these tasks, care should be taken in identifying the parallel execution mode, task or pipeline parallelism that should be used. When considering a case study of computational finance tasks, a benchmark workload of Heston and Black-Scholes-based options implemented using OpenCL and OpenSPL, the benefit of this method of using data centre FPGAs is illustrated. These devices deliver latency performance close to that of workstation grade GPUs, while requiring considerably less energy, resulting in 30\% more floating point operations per Joule of energy consumed.
\end{abstract}




\acmformat{Gordon Inggs, 2016. Algorithmic Trading: A brief, computational finance case study on data centre FPGAs}



\begin{bottomstuff}

Author's addresses: Gordon Inggs, Circuits and Systems Research Group, Department of Electrical and Electronic Engineering, Imperial College London, London, United Kingdom.
\end{bottomstuff}

\maketitle

\section{Introduction}

\subsection{Why FPGAs are coming to a data centre near you}

I argue that Field Programmable Gate Arrays (FPGAs) will be increasingly deployed at scale, i.e. in large data centres and available from Infrastructure-as-a-Service (IaaS) providers, due to recent trends in both computing hardware and software.

Firstly with respect to hardware, the ramifications of the end of Dennard scaling are on-going. While the turn to multicore Central Processing Unit (CPU) architectures has continued power efficiency gains, there has been a significant slowdown in the rate of increase of computational power efficiency since 2000~\cite{KoomeyLaw}. 

As a result, alternative computational architectures such as Graphics Processing Units (GPUs) and FPGAs are proliferating. These architectures provide orders of magnitude better power efficiency than even the latest multicore CPUs. This power efficiency comes by virtue of architectural specialisation on parallel execution. In the GPU case, this is in the form of Single Instruction, Multiple Data (SIMD) parallelism, while FPGAs can provide fine-grained pipeline parallelism by creating custom architectures.

The performance and economies of scale attached to GPUs make them attractive for data centre deployment, and indeed, several IaaS providers offer GPU resources. However, GPUs are typically power \emph{intensive}. So, while offering high throughput performance, this comes at increased power consumption relative to a server grade CPU, and more importantly, increased heat generation which increases the data centre infrastructure costs. FPGAs, by contrast, are much less power intensive, and can often be passively cooled, even when deployed at scale.

Secondly with respect to software, a significant development is the widening pool of programmers that can potentially use FPGAs thanks to the  of High Level Synthesis (HLS) tools. As I discussed in previous work~\cite{FPT2014}, it is now quite possible to produce FPGA implementations that are competitive with other HPC approaches.

Hence, in the past users targeting FPGAs were frequently referred to as designers, as the process of using the devices was much more akin to hardware design. However, with the recent advances in HLS, it is increasingly becoming indistinguishable from programming, albeit with very long compile times. Hence, in this note, I refer to users of FPGAs as programmers.


\subsection{Data Centre FPGAs}

In the previous subsection I argued that it is increasingly attractive to deploy FPGAs en-masse to data centres. IaaS providers could then make these FPGAs available as virtualised compute resources, similar to their current multicore CPU and GPUs offerings. 

I suggest that the availability of FPGAs in the cloud will result in a distinct category of \emph{data centre FPGAs}. Arguably, the offerings from the two largest FPGA vendors, Xilinx and Altera, are already bifurcated between large devices intended for high performance computing, such as the Virtex and Stratix lines of chips, and smaller, embedded application-focused devices, such as the System-on-Chip Zynq and Cyclone lines.

In this note, I consider data centre FPGAs to have the following three characteristics:
\begin{enumerate}
    \item \emph{Size} - being one of the largest devices available for that process technology.
    \item \emph{Hosted} - programmed and controlled by, and as well capable of communicating with a conventional host CPU. A single host CPU may manage multiple data centre FPGAs.
    \item \emph{Scalable} - part of a modular system which can be replicated many times over.
\end{enumerate}

Note, that the characteristics above do not preclude the FPGA from having the capability to communicate with other systems, besides its host CPU.

\subsection{Questions for Data Centre FPGAs}

\label{sec:Questions}

I believe there to be two key research questions for data centre FPGAs.

Firstly, what should the FPGAs be used for, relative to other data centre computing resources such as multicore CPUs and GPUs?

Secondly, how should the FPGA resources be used, given the opportunity for different forms of parallel execution?

Both research questions are the subject of much ongoing research, however in this note I largely focus on the second question.

\subsection{Contributions}

In this note, I make the following three distinct contributions in determining how data centre FPGAs should be used:

\begin{enumerate}
    \item A methodology for using data centre FPGAs, both in terms of what portion of the task should be implemented on the FPGA as well how the work should be implemented upon the device. 
    \item A case study, applying the methodology I outline to the real world problem domain of option pricing from computational finance.
    \item An evaluation of this methodology, using the case study. I describe the efficiency of data centre FPGAs from three leading providers, using two heterogeneous programming standards. I also compare these devices to a multicore Intel CPU and GPUs from both major vendors, AMD and NVIDIA.
\end{enumerate}

\subsection{The rest of the note}

The remainder of this note describes my suggested approach to using data centre FPGAs, and a preliminary evaluation of the approach that I outline. In the next section, I describe the approach.

\section{Using data centre FPGAs}

\subsection{What data centre FPGAs should be used for}

I propose the following method for using data centre FPGAs, derived from \cite{HC_goals_methods_Open_Problems}'s model for using heterogeneous computing systems, as illustrated in Figure~\ref{fig:FPGADataCentreFlowchart}:

\begin{figure}
\centering
\includegraphics[width=0.8\textwidth]{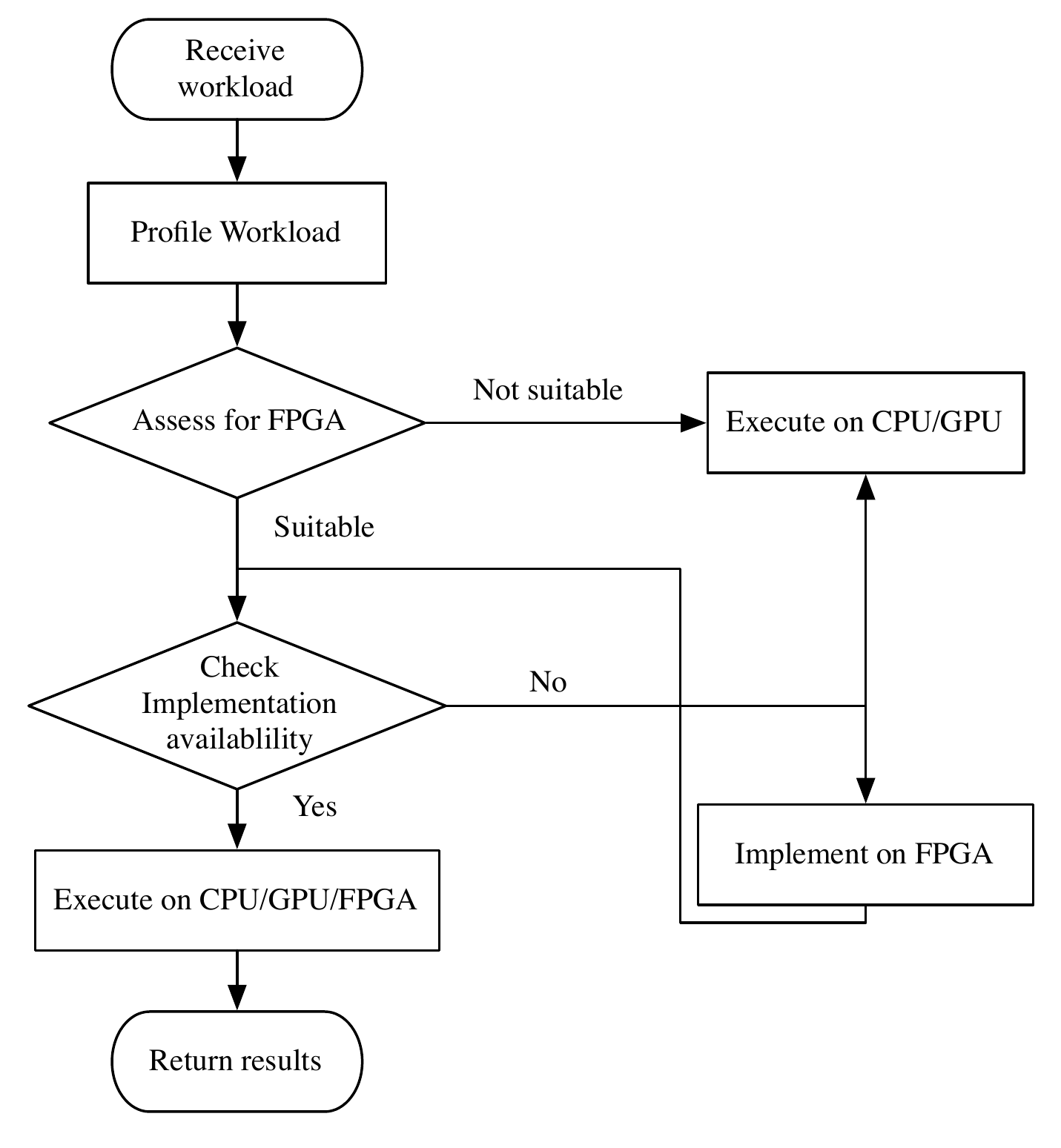}
\caption{Proposed data centre FPGA method.}
\label{fig:FPGADataCentreFlowchart}
\end{figure}

\begin{itemize}
    \item \emph{Receive Workload}: the workload of computational tasks are specified by the programmer, possibly using a general purpose programming language.
    \item \emph{Profile Workload}: the computational tasks are analysed with respect to the different parallel execution modes. This could either be done by running a subset of the tasks, or a static analysis comparing the tasks to heterogeneous computing benchmarks~\cite{AutomaticPortableMappingOpenCL}, such as Rodinia~\cite{Rodinia}.
    \item \emph{Assess for FPGA}: based upon the workload profiling, an assesment needs to be performed to decide whether FPGAs should be employed, and if so, for which tasks or proportion of the tasks should be executed on the FPGAs.
    \item \emph{Execute on CPU/GPU}: if the workload is found to not be suitable for FPGA execution, or if it needs to be implemented upon the FPGA, some of the task can be implemented upon available CPU or GPU resources, provided doing so doesn't exceed programmer constraints such as cost.
    \item \emph{Check implementation availability}: if FPGAs are being used, implementations of the task for FPGAs need to be found. 
    \item \emph{Implement on FPGA}: if there is no FPGA implementation available, one needs to be generated. The implementation should seek to achieve the programmers objectives such as latency minimisation as efficiently as possible.
    \item \emph{Execute on CPU/GPU/FPGA}: when the workload executing on FPGAs, other available resources should also be used, using the complimentary strengths of all, as I have described in other work~\cite{FSP2015}.
    \item \emph{Return results}: the results are returned, in the form expected by the programmer.
\end{itemize}

The degree of automation in this method is left up to the discretion of the system implementer.

\subsection{How to make the most of data centre FPGAs}

In the method outlined above, there are many open questions, such as the best method for profiling workloads or how to partition tasks between heterogeneous computing platforms, or even the overall question how of such a method should be abstracted to programmers.

In this note however, as noted in Section~\ref{sec:Questions}, I am primarily concerned with proposing this method, and considering the implementation of tasks upon FPGAs. 

As demonstrated in my previous work~\cite{FPT2014}, it is increasingly possible to use HLS flows to implement tasks upon FPGAs, using open standards such as OpenCL~\cite{opencl} and OpenSPL~\cite{openspl}. However, these standards only guarantee functional correctness, leaving any optimisation of the platform up to the programmer.

However, due to the inherent flexibility of FPGAs, the programmer is still faced with a choice between different forms of parallelism. I suggest that the programmer should evaluate many possible architectures, and select that which offers the highest power efficiency, i.e. the most computational effort for the energy expended.

\section{Derivatives Pricing Case Study}

\subsection{Background}

\subsubsection{Option Pricing}
Computational finance is an important activity in modern commerce. The problems in the area are concerned with the modelling of uncertainty or risk. Derivatives pricing is one of the largest activities in this area, with $\approx\$100$ trillion of derivatives products currently active. Derivative pricing is also computationally intensive, and as a result is a major consumer of high performance computing, including multicore CPUs, GPUs and increasingly, FPGAs.

An example of a derivative is an \emph{option} contract. An option is an agreement where a holder pays a \emph{premium} to the writer in order to obtain rights with regards to an \emph{underlying}, an asset such as a stock or commodity. This right either allows the holder to buy or sell the underlying at a defined \emph{strike price} at a defined \emph{exercise time}. 

The holder has bought the \emph{right} to exercise the transaction if they so choose, and is in no way obligated to so. In derivatives pricing, the intrinsic value of the option is the \emph{payoff}, the difference between the strike price and \emph{spot price} of the underlying at the exercise time, or zero, whichever is higher~\cite{Hull}.


\subsubsection{Monte Carlo-based Option Pricing}
\begin{figure}
\centering
\includegraphics[width=0.8\textwidth]{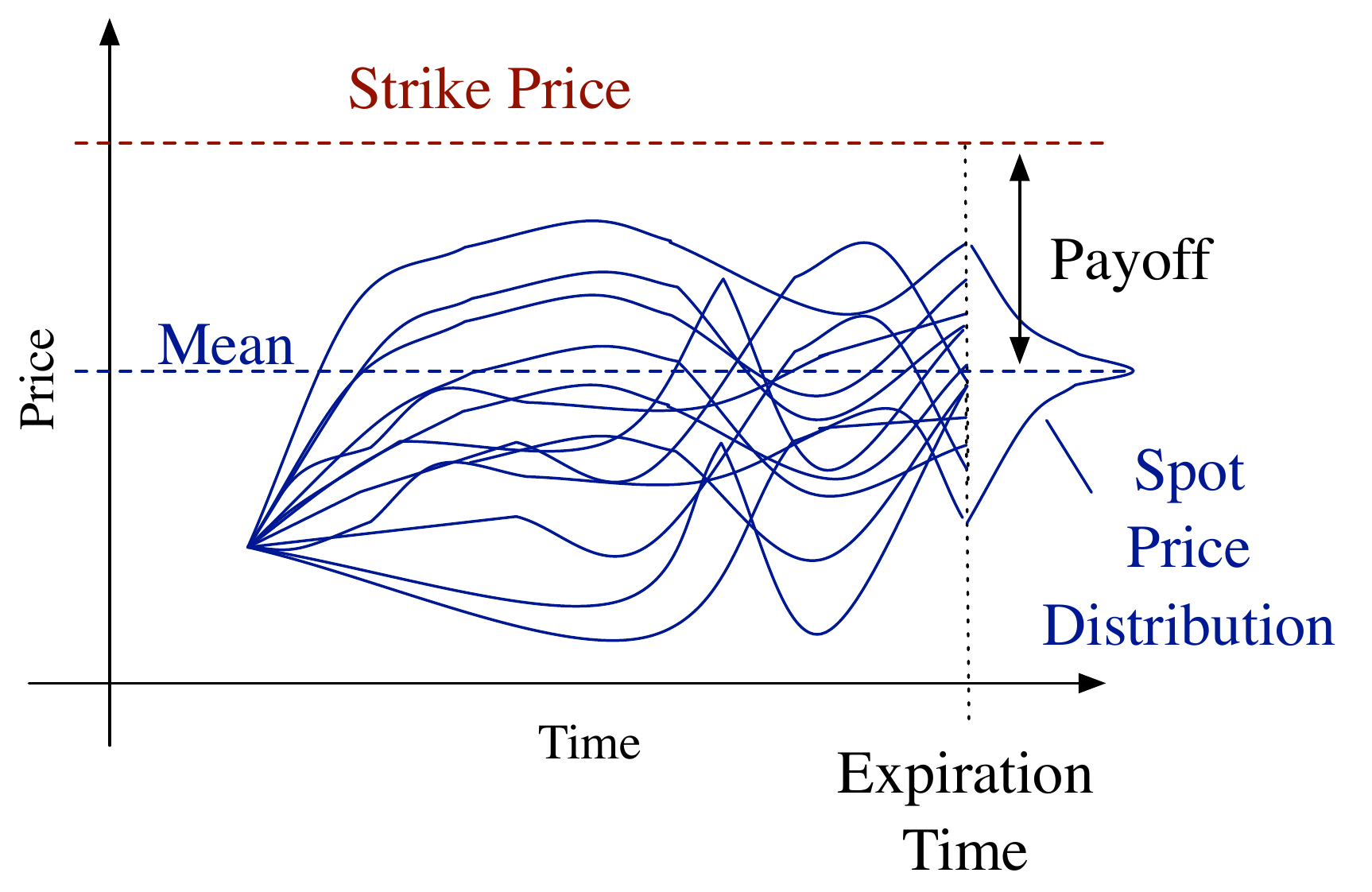}
\caption{Overview of Monte Carlo derivatives pricing.}
\label{fig:OptionMCOverview}
\end{figure}

The popular Monte Carlo technique for option pricing uses random numbers to create scenarios or simulation paths for the underlying based upon a model of its spot price evolution. The average outcome of these paths is then used to approximate the payoff~\cite{Hull}, i.e. $${V_{t}=e^{-r(T-t)}\int_{w}V(w)d\mathbf{P}(w)\approx e^{-r(T-t)}\frac{1}{N}\sum_{i=0}^{N-1}V(S_{i})},$$

where $V_{t}$ is the current value of the option, $e^{-r(T-t)}$ the discount factor, $P(w)$ the probability space defined by the underlying asset and $S_{i}$ the price of the asset. I have provided an illustration of Monte Carlo option pricing in Figure \ref{fig:OptionMCOverview}. 

Although computational expensive, the Monte Carlo pricing technique is robust, capable of tolerating underlying models with many more stochastic variables than competing methods~\cite{Hull}.

\subsubsection{Computational Implementation}

The Monte Carlo option pricing algorithm can be expressed using the MapReduce computational design pattern~\cite{MapReduce}. The simulation paths comprise the map operation, while the reduction is the average of the payoffs. I have described this pattern in C code in Listing~\ref{lst:MCMapReduce}, with MAP and REDUCE labels.

\begin{lstlisting}[language=C,caption={Monte Carlo Option Pricing as MapReduce},basicstyle=\ttfamily,label=lst:MCMapReduce]
MAP:
    for(i=0;i<PATHS;++i){
        state = path_init(seed++);
        for(j=0;j<PATH_POINTS;++j) state = path(state);
        value[i] = payoff(state);
    } 

REDUCE:
    for(i=0;i<PATHS;++i) result += value[i]/PATHS;
\end{lstlisting}

This expression of the algorithm highlights another advantage of the Monte Carlo approach - it is extremely amenable to parallel execution, as each simulation, i.e. each iteration of the outer loop in the MAP section, can be performed in parallel. In fact, it is considered to the canonical ``Embarrassingly Parallel'' algorithm~\cite{view_from_berkeley_2006}.

\subsection{Data centre FPGA Derivatives Pricing}

The C description in Listing~\ref{lst:MCMapReduce} is useful when deciding how Monte Carlo derivatives pricing should be mapped to a data centre FPGA. 

The majority of the computational work is clearly in the \emph{MAP} code section. In particular, the path function call requires the generation of multiple Gaussian random numbers. Hence, this suggests that the FPGA should be specialised on this.

A second consideration is minimising communication between the data centre FPGA and the host CPU. Between the \emph{MAP} and \emph{REDUCE} sections, only the resulting value of each simulation need to be communicated. However, if the code is segmented in any other way, considerably more communication would be required. Hence, this suggests that the FPGA should be used for the \emph{MAP}, while the host CPU should be used for the \emph{REDUCE} code section.

\subsection{Optimising FPGA Derivatives Pricing}

Having decided that it is the simulation paths or \emph{MAP} code section in Listing~\ref{lst:MCMapReduce} that must be implemented upon the FPGA, the next question is how the task should be implemented upon the FPGA. Below I have described two optimisations that might be applied, in terms of two leading programming standards supported by HLS flows, namely OpenCL and OpenSPL.

\subsubsection{Task Parallelism}

I have illustrated \emph{task parallelism} in code Listing~\ref{lst:MCTP} by introducing a third parallel loop bound by $P$. Each iteration of the outer loop can be performed completely independently.

\begin{lstlisting}[language=C,caption={Identifying Task Parallelism},basicstyle=\ttfamily,label=lst:MCTP]
for(p=0;p<P;++p){
    for(i=0;i<PATHS/P;++i){
        state = path_init(seed++);
        for(j=0;j<PATH_POINTS;++j)
            state = path(state); 
        
        offset = p*PATHS/P; 
        value[offset + i] = payoff(state);
    }
}
\end{lstlisting}

In OpenCL, task parallelism is inherent to the standard. The programmer expresses their tasks as instances or \emph{work-items} of programs or \emph{kernels}. Architecturally, the degree of task parallelism available is captured by the number of \emph{compute units} available in the \emph{device}. Hence, the Monte Carlo code in Listing~\ref{lst:MCTP} could be used, without the outer, parallel loop, and the number of work-items set to $P$.

In OpenSPL, task parallelism is expressed using an architectural loop, which creates multiple copies of loop body. Hence code very similar to Listing~\ref{lst:MCTP} must be used, with the outer loop being the architectural description.

\subsubsection{Pipeline Parallelism}

In the naive formulation of the code, there is already ample opportunity for pipeline parallelism, as each computational operation could be viewed as a stage in the pipeline. However, the iterations of inner loop bounded by PATH\_POINTS are data dependent, and hence would stall any pipeline generated in this naive fashion.

Hence, in order to further extend \emph{pipeline parallelism}, I have unrolled the inner loop in my code, as demonstrated in Listing~\ref{lst:MCPP}.

\begin{lstlisting}[language=C,caption={Doubling the potential Pipeline Parallelism},basicstyle=\ttfamily,label=lst:MCPP]
for(i=0;i<PATHS;++i){
    state = path_init(seed++);
    for(j=0;j<PATH_POINTS/2;j+=2){
        state2 = path(state);
        state = path(state2);
    }
    value[i] = payoff(state);
}
\end{lstlisting}

The core OpenCL standard doesn't provide a means to express pipeline parallelism, beyond the manual unrolling performed in Listing~\ref{lst:MCPP}. However vendors such as Altera provide code source code pragmas which allow for loop unrolling.

Similarly, OpenSPL also doesn't provide a native means to express pipeline parallelism, beyond the manual form described in Listing~\ref{lst:MCPP}.

\section{Evaluation}

\subsection{Experimental Setup}

In this subsection, I describe how the experimental platforms used, the tasks used in the evaluation, and finally how the results measured were captured.

\subsubsection{Experimental Platforms}

An overview of the experimental platforms used is given in Table~\ref{tab:ExperimentalPlatforms}, with the details of the FPGAs devices given in Table~\ref{tab:ExperimentalFPGAs}.

\begin{table} 
\caption{Configuration of Experimental Data Centre FPGA Platforms} 
\addtolength{\tabcolsep}{-2pt}
\begin{tabular}{cccc}
\label{tab:ExperimentalPlatforms}
Designation & \parbox[c]{4cm}{\centering\center Name \\(FPGA)} & \parbox[c]{2cm}{\centering\center Communication Technology} & \parbox[c]{5cm}{\centering\center Programming Standard \\(tool)}\\ 
\noalign{\smallskip} 
\hline 
\noalign{\smallskip}
P385-D5 & \parbox[c]{4cm}{\center Nallatech P385-D5 \\(Stratix V 5SGSD5)}  & \parbox[c]{2cm}{\center PCIe} & \multirow{2}{*}{\parbox[c]{5cm}{\centering\center OpenCL \\(Altera OpenCL SDK 15.0)}} \\

C5-SoC & \parbox[c]{4cm}{\center Altera Cyclone 5 SoC \\(Cyclone V 5CSXC6)}  & \parbox[c]{2cm}{\center AXI} & \\

\noalign{\smallskip}
\hline
\noalign{\smallskip}

Max3 & \parbox[c]{4cm}{\center Maxeler Max 3424A \\(Virtex 6 XC6VSX475T)} & \parbox[c]{2cm}{\center PCIe} & \multirow{2}{*}{\parbox[c]{5cm}{\centering\center OpenSPL \\(Maxeler MaxCompiler 14.1)}} \\

Max4 & \parbox[c]{4cm}{\center Maxeler Max 4 \\(Stratix V 5SGSD8)}  & \parbox[c]{2cm}{\center Ethernet} & \\

\noalign{\smallskip}

\hline 
\end{tabular} 
\end{table}

\begin{table}
\caption{Experimental FPGA Resources}
\addtolength{\tabcolsep}{-2pt}
\begin{tabular}{cccccccc}
\label{tab:ExperimentalFPGAs}
\parbox[c]{3.75cm}{\center\centering FPGA} & \parbox[c]{1.5cm}{\center\centering Vendor} & \parbox[c]{0.8cm}{\center\centering CMOS Size (nm)} & \parbox[c]{1.5cm}{\center\centering Targeted Clockrate (MHz)} & \parbox[c]{1.5cm}{\center\centering LookUp Tables (LUTs)} & \parbox[c]{1.75cm}{\center\centering Flipflop Registers (FFs)}& \parbox[c]{1cm}{\center\centering Block RAMs (BRAMs)} & \parbox[c]{1cm}{\center\centering DSPs}\\ 
\noalign{\smallskip} 
\hline 
\noalign{\smallskip}
\parbox[c]{3.75cm}{Stratix V 5SGSD5} & \parbox[c]{1.5cm}{Altera} & 28 & 250 & 457k & 690k & 2014 & 1590\\ 

\parbox[c]{3.75cm}{Cyclone V 5CSXC6} & \parbox[c]{1.5cm}{Altera} & 28 & 250 & 110k & 41.51k & 557 & 112\\ 

\noalign{\smallskip}
\hline
\noalign{\smallskip}

\parbox[c]{3.75cm}{Virtex 6 XC6VSX475T} & \parbox[c]{1.5cm}{Xilinx} & 40 & 200 & 298k & 595k & 1064 & 2016\\ 

\parbox[c]{3.75cm}{Stratix V 5SGSD8} & \parbox[c]{1.5cm}{Altera} & 28 & 180 & 695k & 1050k & 2567 & 1963\\ 

\noalign{\smallskip}
\hline 
\end{tabular} 
\end{table}

The three references platforms used are detailed in Table~\ref{tab:ReferencePlatforms}.

\begin{table} 
\caption{Comparison of Reference Platforms} 
\label{tab:ReferencePlatforms}
\begin{tabular}{cccccc} 
\hline\noalign{\smallskip} 
\parbox[c]{4cm}{\center Platforms} & \parbox[c]{1.2cm}{\center CMOS Size\\(nm)} & \parbox[c]{2cm}{\center Clockrate\\(GHz)} & \parbox[c]{1.2cm}{\center Memory\\(GBs)}& \parbox[c]{0.8cm}{\center Threads}& \parbox[c]{4.5cm}{\center Tool}\\ 
\noalign{\smallskip} 
\hline 
\noalign{\smallskip}
\parbox[c]{4cm}{Intel Core i7-2600S} & 32 & 2.8\ & 16 & 8 & GCC 4.8\\
\parbox[c]{4cm}{AMD Firepro W5000} & 28 & 0.825 & 2 & 768 & \parbox[c]{4.5cm}{AMD OpenCL SDK 2.9}\\
\parbox[c]{4cm}{NVIDIA Quadro K4000} & 28 & 0.81 & 3 & 768 & \parbox[c]{4.5cm}{NVIDIA OpenCL SDK 7.0}\\
\hline \end{tabular} 
\end{table}

\subsubsection{Option Pricing Tasks}

An overview of the 5 option pricing tasks are given Table~\ref{tab:OptionTasks}. These tasks are drawn from the Kaiserslautern Option Pricing benchmark~\cite{KS_FPGA_Paper}, as well as the work from Imperial College London on pricing Black-Scholes Model-based Asian options~\cite{Asian_Option_Paper}.

\begin{table}
\centering
\caption{Overview of Experimental Option Pricing Tasks} 
\begin{tabular}{cccc}
\label{tab:OptionTasks}

Designation & Underlying & Option & \parbox[c]{3cm}{\centering Complexity \\($\frac{\text{FLOP}}{\text{Simulation}}$)} \\

\noalign{\smallskip}
\hline
\noalign{\smallskip}

he-eu & Heston & European & 323590 \\
he-ba & Heston & Barrier & 327686 \\
he-do & Heston & Double Barrier & 331780 \\
he-di & Heston & Digital Double Barri & 331781 \\
bl-as & Black Scholes & Asian & 147462 \\

\end{tabular} 
\end{table}

Each task was performed with 10 million simulation paths, with 4096 path points in each path.

\subsubsection{Result Measurement}

Two metrics were used in this study:

\begin{itemize}
    \item \emph{Latency}: the latency reported is wall-clock time, i.e. the absolute time passed from task initialisation to the result being returned, as reported by an external time reference. In all cases the system time is used, which is set using the Network Time Protocol.
    \item \emph{Energy}: the energy figures reported are based upon total system power, as measured using an Olson inline power meter. The power meter was polled regularly, with the time since the last measurement used to calculate the energy consumed in the interval.
\end{itemize}

\subsection{Experimental Results}

The latency results for the experimental tasks given in Table~\ref{tab:OptionTasks}, run upon experimental platforms, described in Tables~\ref{tab:ExperimentalPlatforms} and \ref{tab:ReferencePlatforms}, are given in Table~\ref{tab:ExperimentalLatencyResults}. The energy results are given in Table~\ref{tab:ExperimentalEnergyResults}. 

The resource use for FPGA platforms in Table~\ref{tab:ExperimentalPlatforms} are given in Table~\ref{tab:ExperimentalLUTResults}.

Results of note are the relative high base power consumption of the Maxeler Max4 platform, which is approximately 240W when idling, compared to the 69W idling power of the other platforms. Another consideration is relatively small size of the Altera C5-SoC FPGA, which did not allow for any optimisations to be applied.

\begin{table}
\caption{Full Experimental Latency Results in seconds. Platform and Task designations are given in Tables~\ref{tab:ExperimentalPlatforms} and \ref{tab:OptionTasks}. \emph{base} refers to the baseline implementations, \emph{tp} and \emph{pp} to a task and pipeline parallel implementations. The lowest values for each task are given in \textbf{bold}.}

\addtolength{\tabcolsep}{-4pt} 
\begin{tabular}{c|ccc|ccc|ccc|ccc|ccc|}
\centering
\label{tab:ExperimentalLatencyResults}
\multirow{2}{*}{Platform} & \multicolumn{3}{c|}{he-eu} & \multicolumn{3}{c|}{he-ba} & \multicolumn{3}{c|}{he-do} & \multicolumn{3}{c|}{he-di} & \multicolumn{3}{c|}{bl-as} \\
& base & tp & pp & b & tp & pp & base & tp & pp & base & tp & pp & base & tp & pp \\

\noalign{\smallskip}
\hline
\noalign{\smallskip}

AoC P385-D5 & 180 & 24 & \textbf{20} & 178 & 27 & \textbf{22} & 176 & 26 & \textbf{22} & 169 & 26 & \textbf{22} & 189 & 13 & \textbf{11} \\

AoC C5-SoC & 393 & - & - & 347 & - & - & 347 & - & - & 370 & - & - & 343 & - & - \\

Maxeler Max3 & 212 & \textbf{25} & 32 & 213 & \textbf{25} & 30 & 213 & \textbf{24} & 30 & 213 & \textbf{25} & 30 & 217 & \textbf{12} & 23 \\

Maxeler Max4 & 235 & \textbf{24} & 43 & 236 & \textbf{22} & 43 & 236 & \textbf{20} & 43 & 236 & \textbf{24} & 43 & 243 & \textbf{13} & 49 \\

\noalign{\smallskip}
\hline
\noalign{\smallskip}

Intel i7-2600S & - & 1295 & - & - & 611 & - & - & 716 & - & - & 742 & - & - & 517 & - \\

AMD W5000 & - & 8 & - & - & 161 & - & - & 95 & - & - & 111 & - & - & 6 & - \\

NVIDIA K4000 & - & 14 & - & - & 16 & - & - & 17 & - & - & 17 & - & - & 12 & - \\

\end{tabular} 
\end{table}

\begin{table}
\centering
\caption{Full Experimental Energy Results in kilojoules. Platform and Task designations are given in Tables~\ref{tab:ExperimentalPlatforms} and \ref{tab:OptionTasks}. \emph{base} refers to the baseline implementations, \emph{tp} and \emph{pp} to a task and pipeline parallel implementations. The lowest values for each task are given in \textbf{bold}.}

\addtolength{\tabcolsep}{-4pt} 
\begin{tabular}{c|ccc|ccc|ccc|ccc|ccc|}
\centering
\label{tab:ExperimentalEnergyResults}
\multirow{2}{*}{Platform} & \multicolumn{3}{c|}{he-eu} & \multicolumn{3}{c|}{he-ba} & \multicolumn{3}{c|}{he-do} & \multicolumn{3}{c|}{he-di} & \multicolumn{3}{c|}{bl-as} \\
 & base & tp & pp & b & tp & pp & base & tp & pp & base & tp & pp & base & tp & pp \\

\noalign{\smallskip}
\hline
\noalign{\smallskip}

AoC P385-D5 & 13.1 & 2.0 & \textbf{1.7} & 12.8 & 2.2 & \textbf{1.9} & 12.7 & 2.1 & \textbf{1.9} & 12.2 & 2.2 & \textbf{1.9} & 13.6 & 1.1 & \textbf{0.9} \\

AoC C5-SoC & 6.7 & - & - & 5.9 & - & - & 5.9 & - & - & 6.3 & - & - & 5.6 & - & - \\

Maxeler Max3 & 14.9 & \textbf{1.9} & 2.5 & 14.9 & \textbf{2.0} & 2.4 & 14.8 & \textbf{2.0} & 2.4 & 14.8 & \textbf{2.0} & 2.4 & 15.0 & \textbf{0.9} & 1.8 \\

Maxeler Max4 & 58.3 & \textbf{6.2} & 11.1 & 59.1 & \textbf{5.7} & 10.9 & 58.8 & \textbf{5.3} & 11.0 & 59.1 & \textbf{6.2} & 11.0 & 60.5 & \textbf{3.4} & 12.3 \\

\noalign{\smallskip}
\hline
\noalign{\smallskip}

Intel i7-2600S & - & 74.5 & - & - & 70.5 & - & - & 71.7 & - & - & 70.1 & - & - & 64.2 & - \\

AMD W5000 & - & 0.8 & - & - & 16.1 & - & - & 9.9 & - & - & 11.4 & - & - & 0.6 & - \\

NVIDIA K4000 & - & 1.7 & - & - & 2.2 & - & - & 2.4 & - & - & 2.4 & - & - & 1.6 & - \\

\noalign{\smallskip}
\hline
\noalign{\smallskip}

\end{tabular} 
\end{table}

\begin{table}
\centering
\caption{Full Experimental resource use as percentages. Platform and Task designations are given in Tables~\ref{tab:ExperimentalPlatforms} and \ref{tab:OptionTasks}. \emph{base} refers to the baseline implementations, \emph{tp} and \emph{pp} to a task and pipeline parallel implementations. The highest values for each task are given in \textbf{bold}.}

\addtolength{\tabcolsep}{-4pt} 
\begin{tabular}{c|c|ccc|ccc|ccc|ccc|ccc|}
\label{tab:ExperimentalLUTResults}
\multirow{2}{*}{Platform} & \multirow{2}{*}{Resource} & \multicolumn{3}{c|}{he-eu} & \multicolumn{3}{c|}{he-ba} & \multicolumn{3}{c|}{he-do} & \multicolumn{3}{c|}{he-di} & \multicolumn{3}{c|}{bl-as} \\
 & & base & tp & pp & b & tp & pp & base & tp & pp & base & tp & pp & base & tp & pp \\

\noalign{\smallskip}
\hline
\noalign{\smallskip}

\multirow{4}{*}{AoC P385-D5} & LUT & 22 & \textbf{58} & 49 & 23 & \textbf{62} & 60 & 23 & \textbf{62} & 61 & 23 & \textbf{63} & 61 & 20 & \textbf{72} & 53 \\
& FF & 15 & \textbf{45} & 39 & 16 & \textbf{47} & 46 & 16 & \textbf{48} & 46 & 16 & \textbf{48} & 46 & 13 & \textbf{53} & 36 \\
& BRAM & 25 & \textbf{54} & 44 & 25 & \textbf{54} & 51 & 25 & \textbf{54} & 52 & 25 & \textbf{54} & 52 & 23 & \textbf{67} & 37 \\
& DSP & 6 & \textbf{34} & 27 & 6 & 37 & \textbf{46} & 6 & 37 & \textbf{46} & 6 & 37 & \textbf{46} & 3 & \textbf{39} & 34 \\

\noalign{\smallskip}
\hline
\noalign{\smallskip}

\multirow{4}{*}{AoC C5-SoC} & LUT & 36 & - & - & 38 & - & - & 38 & - & - & 38 & - & - & 29 & - & - \\
& FF & 24 & - & - & 25 & - & - & 25 & - & - & 25 & - & - & 17 & - & - \\
& BRAM & 36 & - & - & 37 & - & - & 37 & - & - & 37 & - & - & 30 & - & - \\
& DSP & 79 & - & - & 88 & - & - & 88 & - & - & 88 & - & - & 46 & - & - \\

\noalign{\smallskip}
\hline
\noalign{\smallskip}

\multirow{4}{*}{Maxeler Max3} & LUT & 13 & \textbf{79} & 69 & 13 & \textbf{81} & 70 & 13 & \textbf{81} & 71 & 13 & \textbf{81} & 71 & 8 & \textbf{80} & 61 \\
& FF & 13 & \textbf{90} & 75 & 13 & \textbf{91} & 78 & 13 & \textbf{91} & 78 & 13 & \textbf{91} & 78 & 8 & \textbf{91} & 65 \\
& BRAM & 6 & \textbf{50} & 40 & 6 & \textbf{50} & 44 & 6 & \textbf{50} & 44 & 6 & \textbf{50} & 44 & 2 & \textbf{43} & 33 \\
& DSP & 4 & \textbf{16} & 14 & 4 & \textbf{16} & 15 & 4 & \textbf{16} & 15 & 4 & \textbf{16} & 15 & 3 & \textbf{18} & 12 \\

\noalign{\smallskip}
\hline
\noalign{\smallskip}

\multirow{4}{*}{Maxeler Max4} & LUT & 10 & \textbf{82} & 57 & 10 & \textbf{82} & 57 & 10 & \textbf{82} & 57 & 10 & \textbf{82} & 57 & 6 & \textbf{85} & 45 \\
& FF & 10 & \textbf{89} & 57 & 10 & \textbf{90} & 57 & 10 & \textbf{90} & 57 & 10 & \textbf{89} & 57 & 6 & \textbf{84} & 43 \\
& BRAM & 3 & \textbf{36} & 24 & 3 & \textbf{36} & 24 & 3 & \textbf{36} & 24 & 3 & \textbf{36} & 24 & 1 & \textbf{34} & 19 \\
& DSP & 10 & \textbf{91} & 49 & 11 & \textbf{91} & 50 & 11 & \textbf{92} & 50 & 11 & \textbf{91} & 50 & 7 & \textbf{96} & 43 \\

\noalign{\smallskip}
\hline
\noalign{\smallskip}

\end{tabular} 
\end{table}

\subsection{Discussion}

\subsubsection{Using Data Centre FPGAs Efficiently}

\begin{figure}
    \centering
    \begin{subfigure}{0.5\textwidth}
    \centering
        \includegraphics[width=\textwidth]{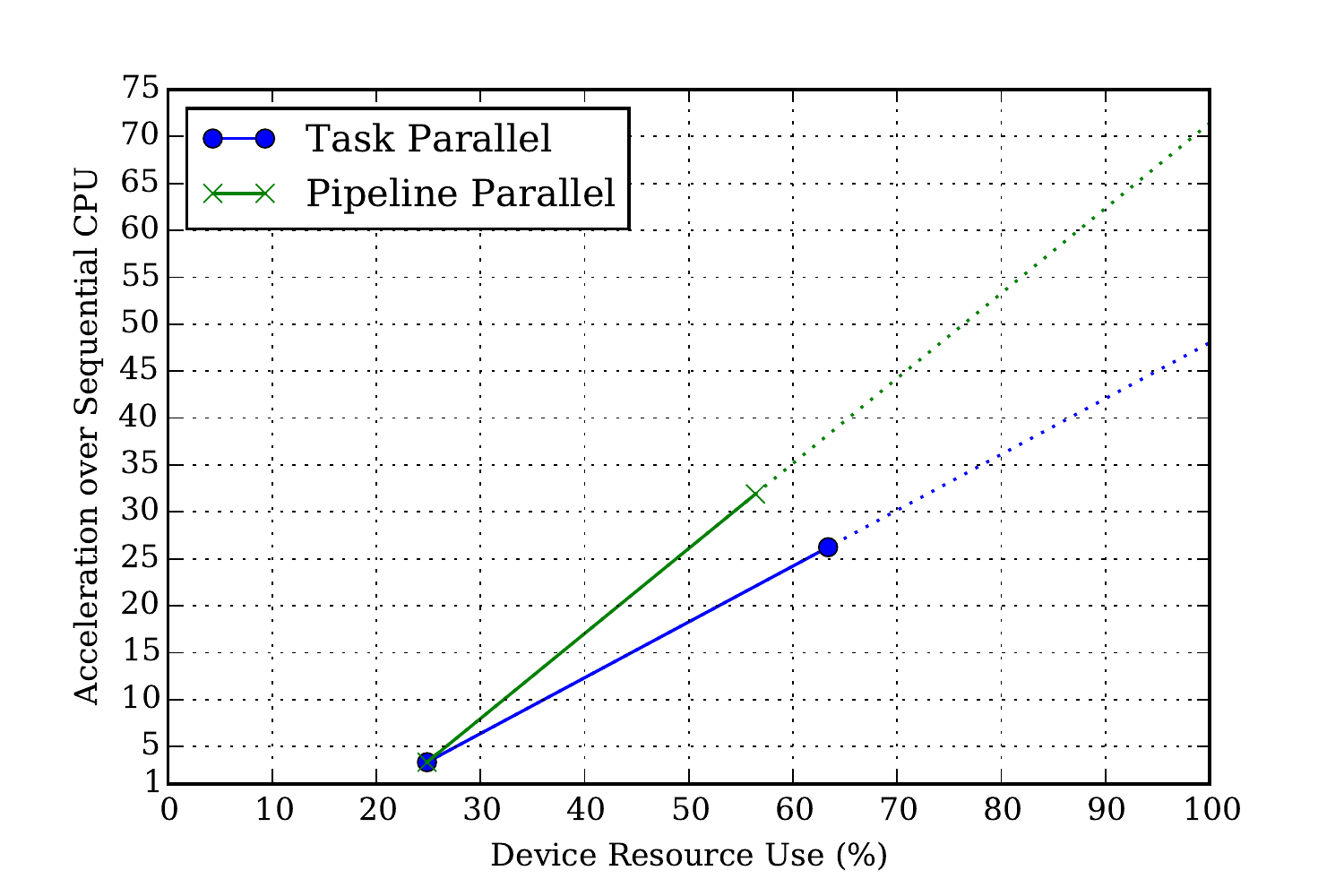}
        \subcaption{Device Use}\label{fig:nallatech_device_use}
    \end{subfigure}%
    \begin{subfigure}{0.5\textwidth}
    \centering
        \includegraphics[width=\textwidth]{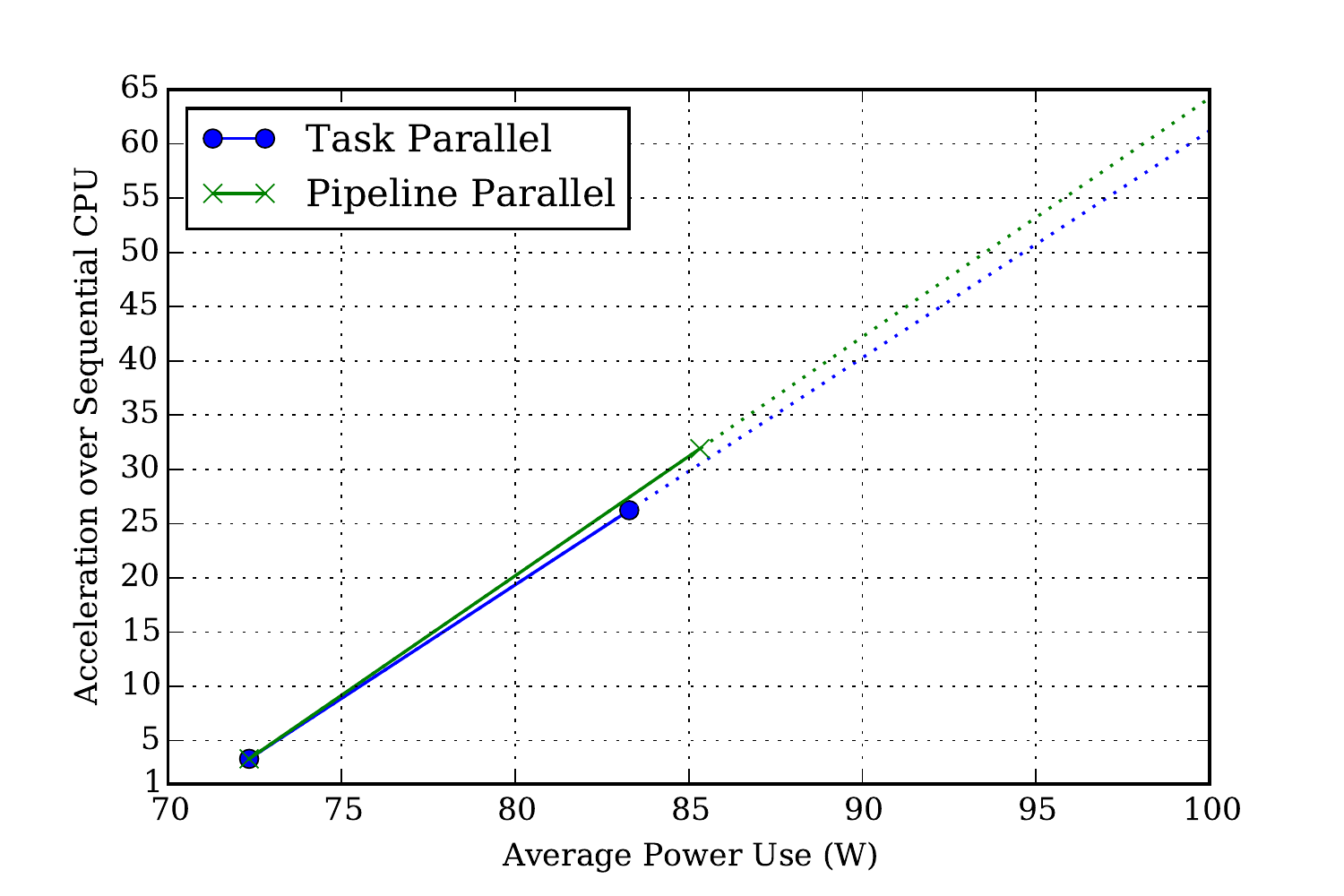}
        \subcaption{Power}\label{fig:nallatech_power_use}
    \end{subfigure}
    \caption{Nallatech P385-D5 Mean Latency Improvement as a function of Device Use and Power.}\label{fig:nallatech_figs}
\end{figure}

\begin{figure}
    \centering
    \begin{subfigure}{0.5\textwidth}
    \centering
        \includegraphics[width=\textwidth]{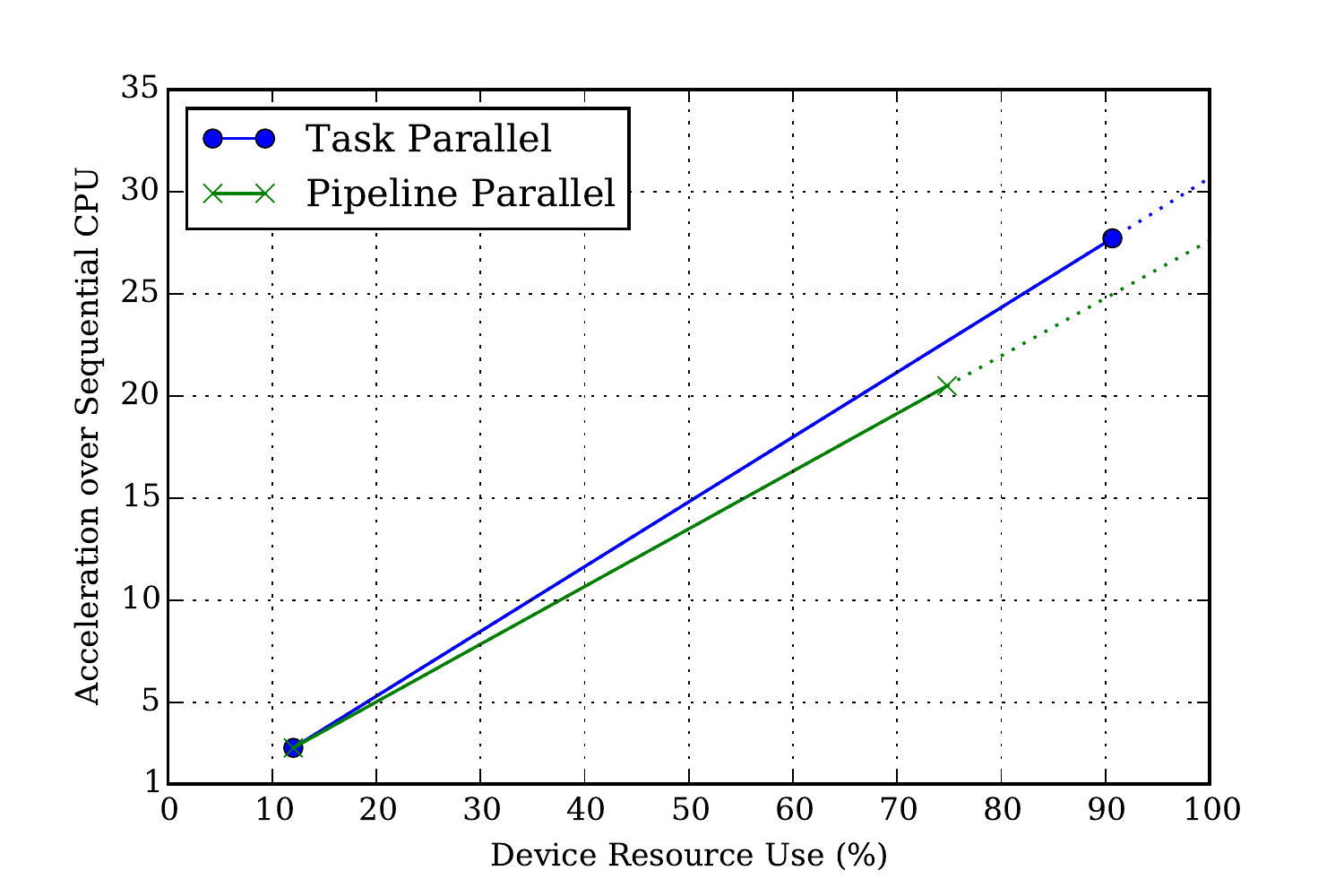}
        \subcaption{Device Use}\label{fig:max3_device_use}
    \end{subfigure}%
    \begin{subfigure}{0.5\textwidth}
    \centering
        \includegraphics[width=\textwidth]{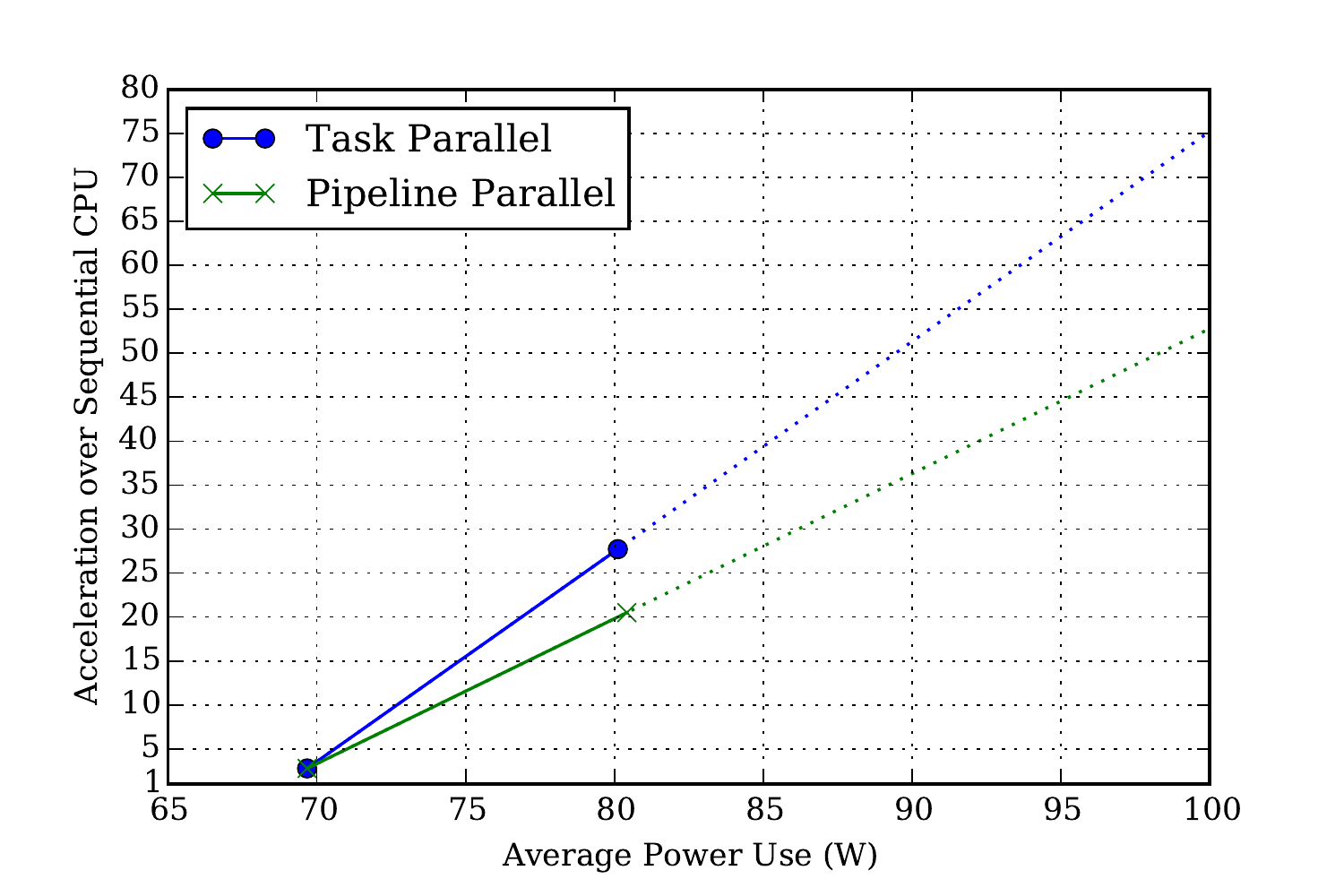}
        \subcaption{Power}\label{fig:max3_power_use}
    \end{subfigure}
    \caption{Maxeler Max3 Mean Latency Improvement as a function of Device Use and Power.}\label{fig:max3_figs}
\end{figure}

\begin{figure}
    \centering
    \begin{subfigure}{0.5\textwidth}
    \centering
        \includegraphics[width=\textwidth]{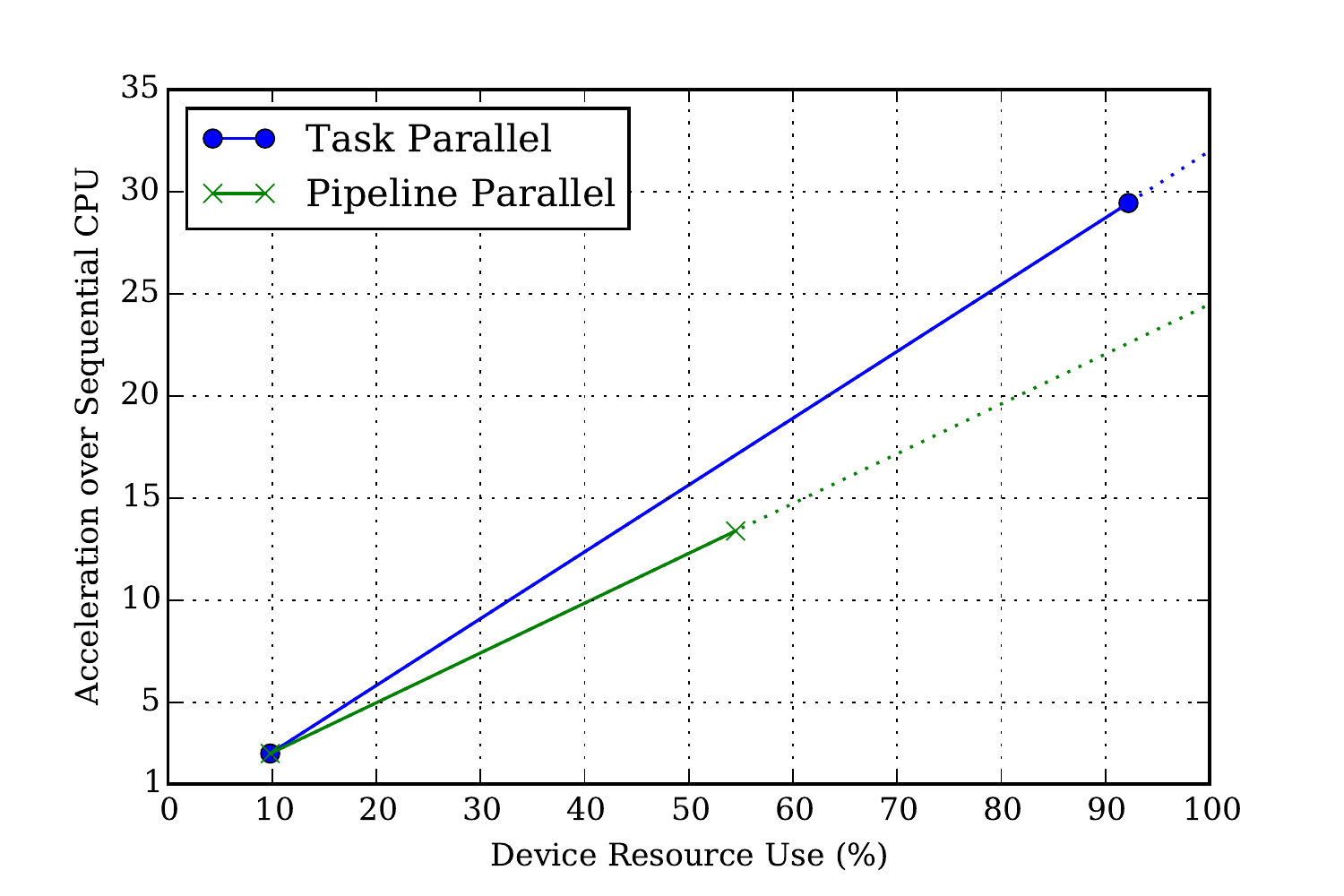}
        \subcaption{Device Use}\label{fig:max4_device_use}
    \end{subfigure}%
    \begin{subfigure}{0.5\textwidth}
    \centering
        \includegraphics[width=\textwidth]{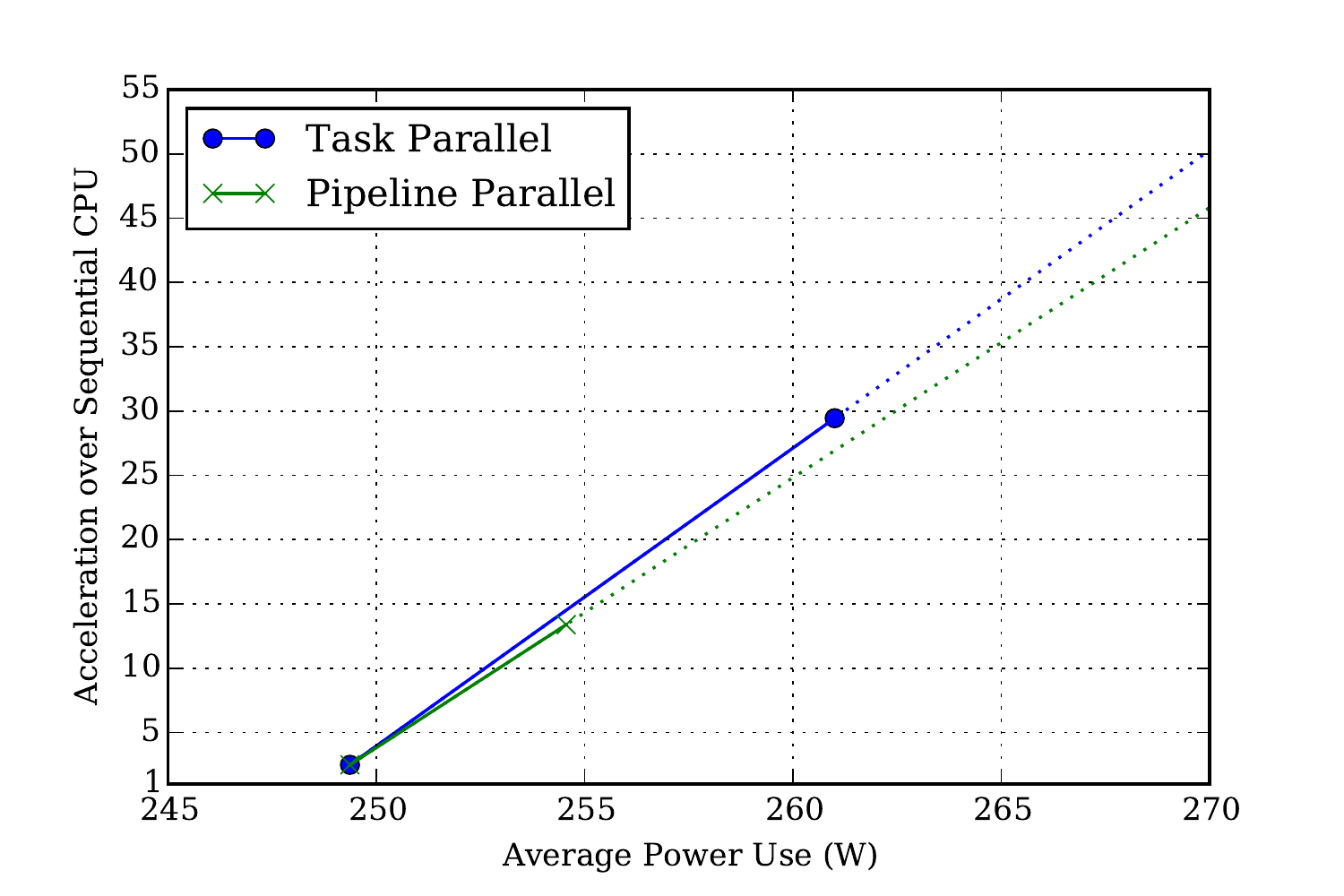}
        \subcaption{Power}\label{fig:max4_power_use}
    \end{subfigure}
    \caption{Maxeler Max4 Mean Latency Improvement as a function of Device Use and Power.}\label{fig:max4_figs}
\end{figure}

In Figures~\ref{fig:nallatech_figs}, \ref{fig:max3_figs} and \ref{fig:max4_figs} I have plotted latency performance for the P385-D5, Max3 and Max4 platforms against a sequential CPU, as a function of power and device resource use for both the task and pipeline parallelism optimisations. I haven't plotted the C5-SoC platform, as no optimisation could be supported on the platform.

In all cases, the baseline implementations show some improvement on the sequential CPU, despite the clockrate of the FPGAs being an order of magnitude less than the reference platform. I posit that this improvement is due to the inherent opportunity for pipelining in the application task, as well as the improvement due to the specialisation of the FPGA architecture.

The results for Nallatech P385-D5 platform, given in Figure~\ref{fig:nallatech_figs}, which uses the Altera OpenCL SDK shows that the pipeline parallelism optimisation results in a more efficient, improvement over the sequential implementation. This is unexpected, given the apparently task parallel nature of the OpenCL standard.

However, the Altera OpenCL SDK transforms the task parallelism of OpenCL into pipeline parallelism, hence any improvement in the opportunity for pipeline parallelism is realised. By contrast, task parallelism replicates compute units, apparently adding additional overhead.

The results for Maxeler platforms, given in Figures~\ref{fig:max3_figs} and \ref{fig:max4_figs}, which use the OpenSPL standard, show that task parallelism optimisations result in a more efficient use of the device and power. Although the benefit over pipeline parallelism optimisations is less pronounced than in the P385-D5 case. 

Similar to the P385-D5, this result is somewhat unexpected, given the inherent orientation of the OpenSPL standard to dataflow architectures, and hence pipeline parallelism. I believe this is due to the increased potential for resource use within pipeline stages that increasing task parallelism allows.

\subsubsection{Using Data Centres FPGAs}

\begin{figure}
    \centering
    \begin{subfigure}{0.5\textwidth}
    \centering
        \includegraphics[width=\textwidth]{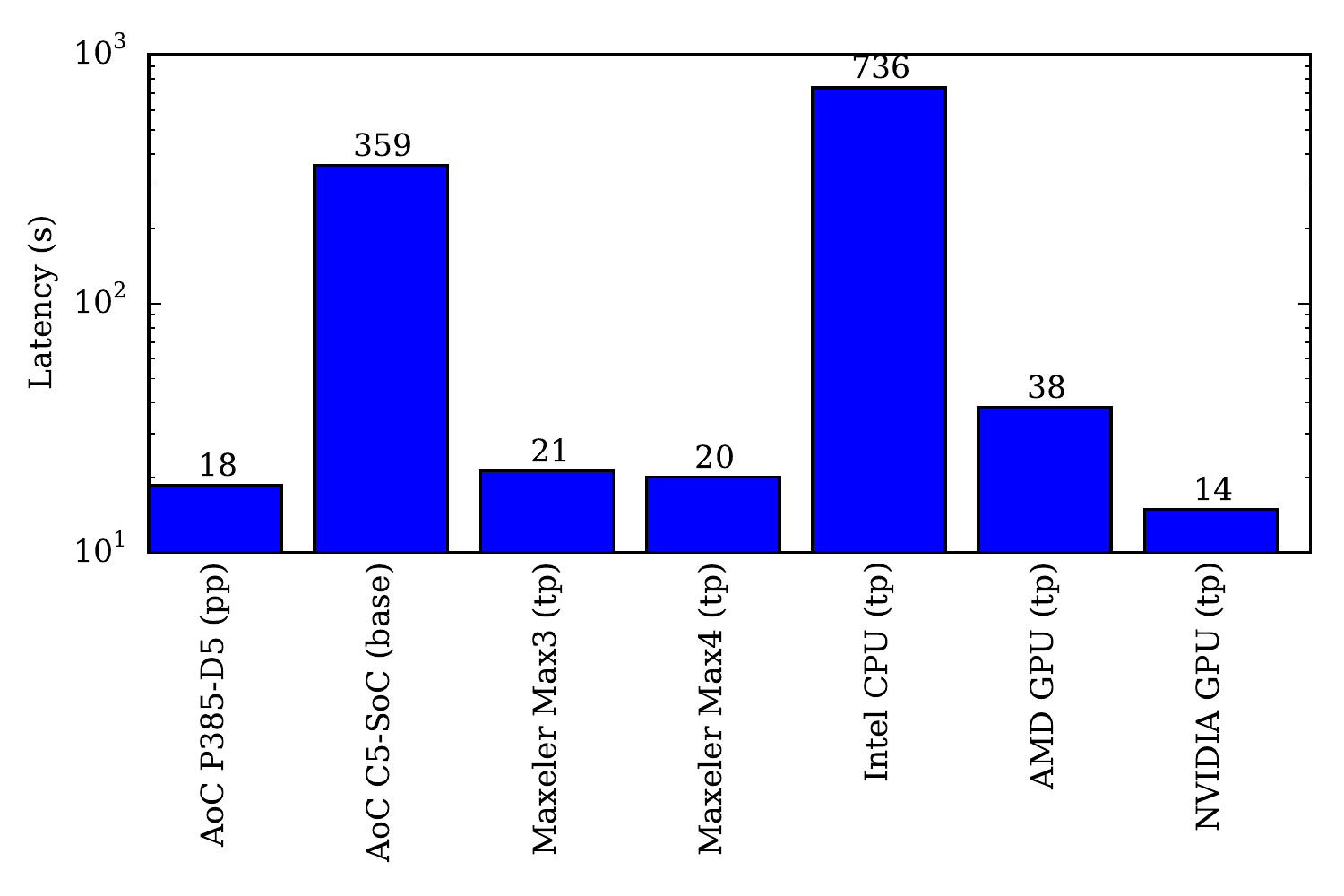}
        \subcaption{Mean Latency}\label{fig:latency}
    \end{subfigure}%
    \begin{subfigure}{0.5\textwidth}
    \centering
        \includegraphics[width=\textwidth]{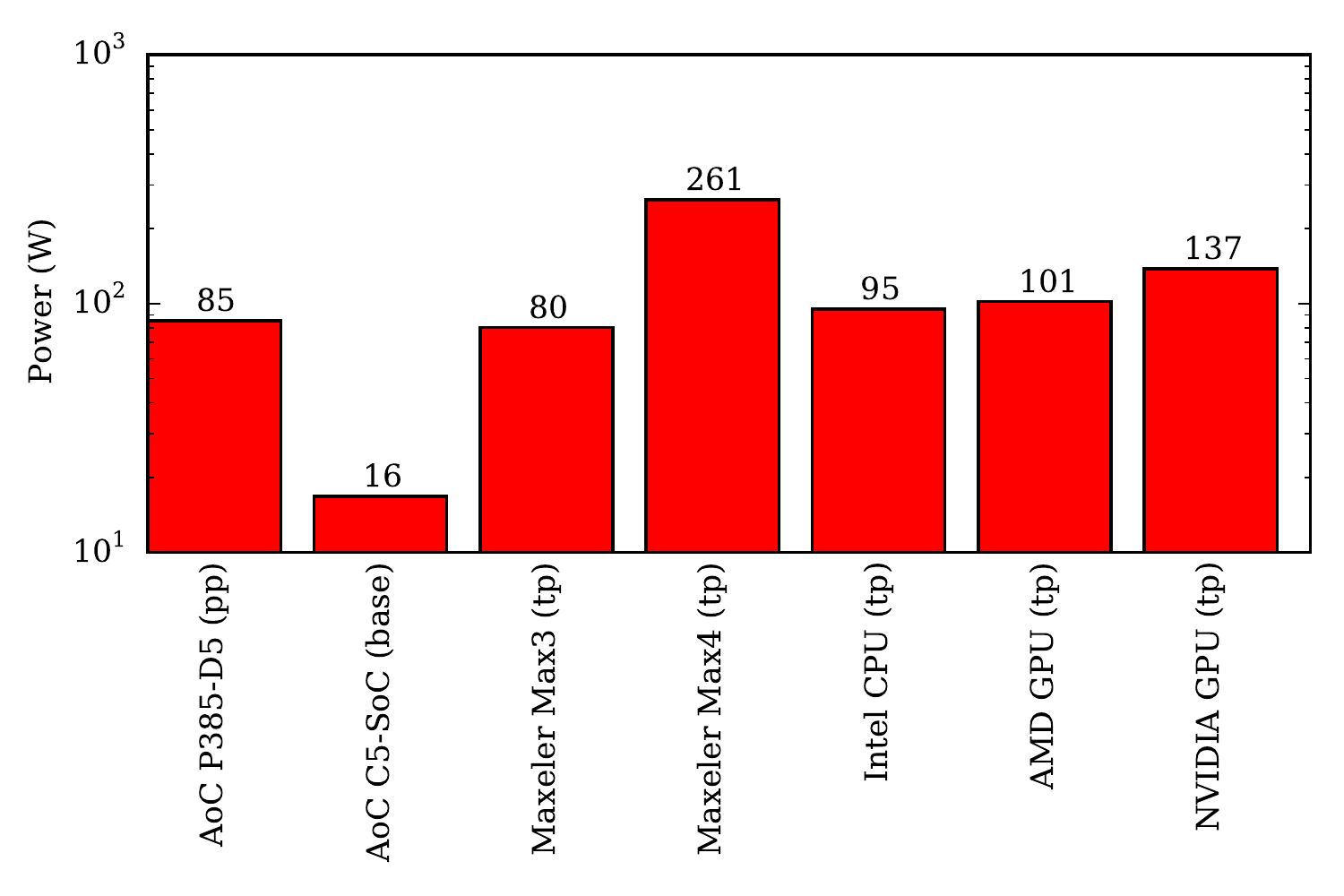}
        \subcaption{Mean Average Power}\label{fig:power}
    \end{subfigure}
    
    \begin{subfigure}{0.5\textwidth}
    \centering
        \includegraphics[width=\textwidth]{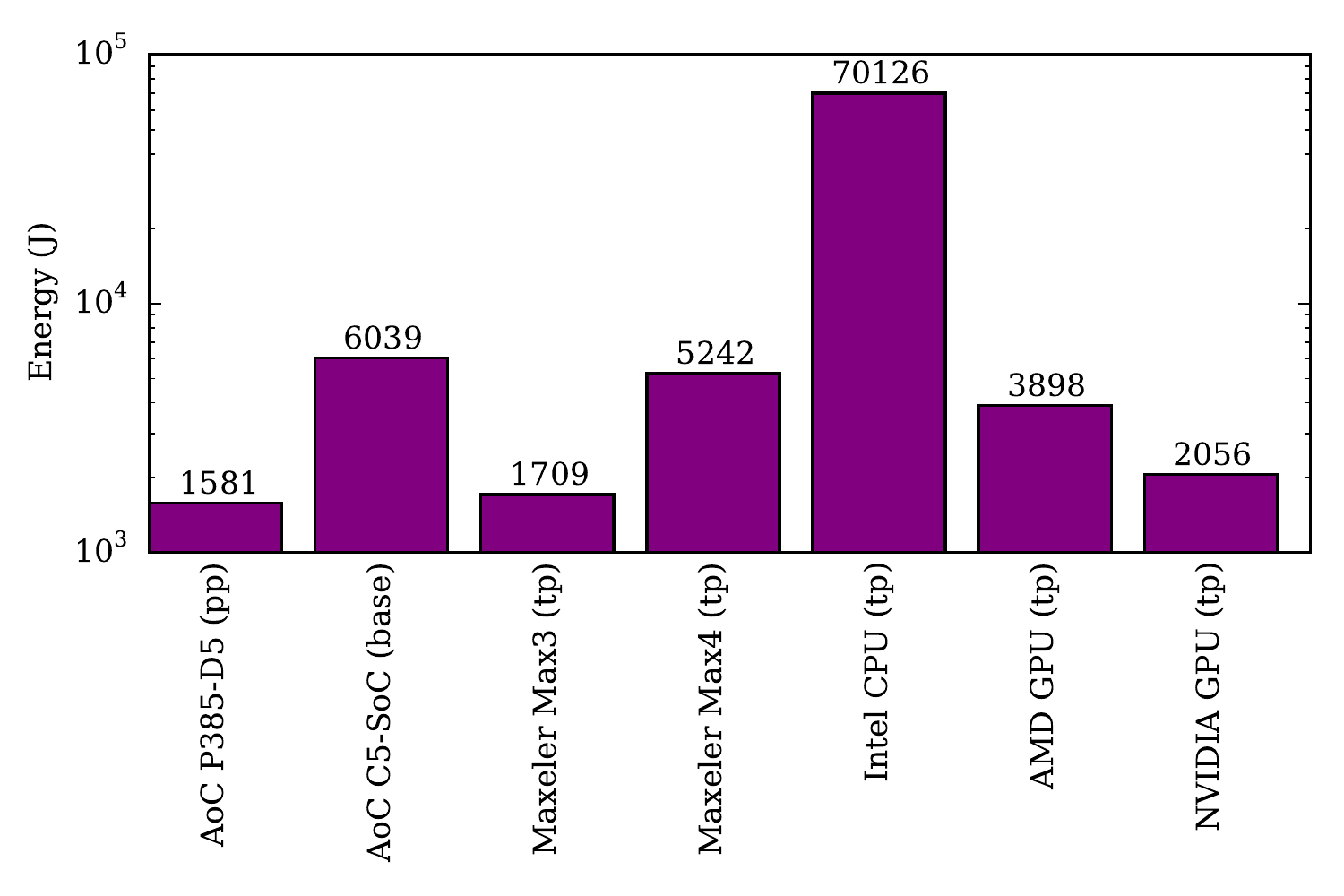}
        \subcaption{Mean Energy}\label{fig:energy}
    \end{subfigure}%
    \begin{subfigure}{0.5\textwidth}
    \centering
        \includegraphics[width=\textwidth]{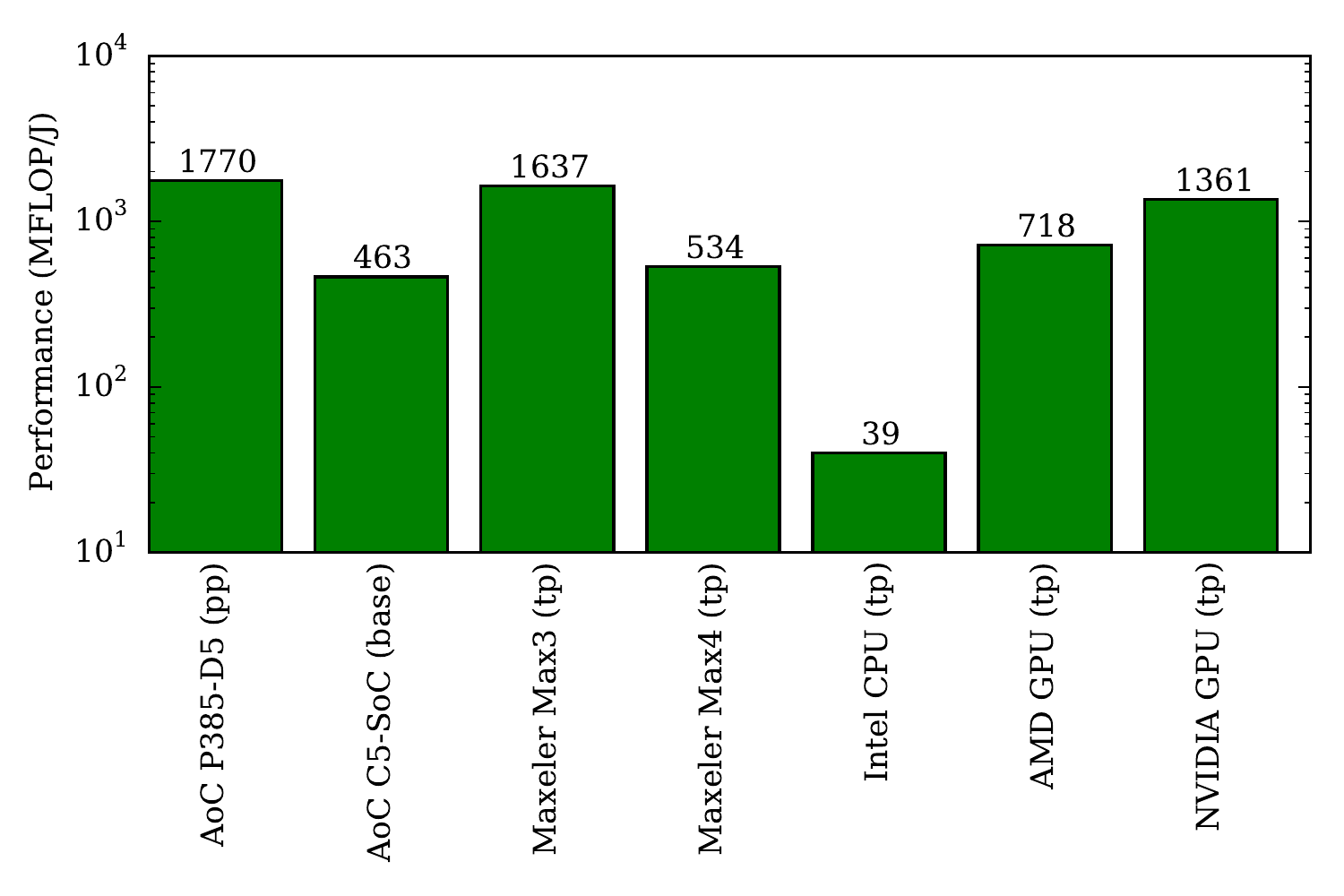}
        \subcaption{Mean Performance Efficiency}\label{fig:performance}
    \end{subfigure}
    \caption{Comparison of platforms in Tables~\ref{tab:ExperimentalPlatforms} and \ref{tab:ReferencePlatforms}, running the tasks in Table~\ref{tab:OptionTasks}.}\label{fig:platform_comparisons}
\end{figure}

Figure~\ref{fig:platform_comparisons} compares all of the experimental platforms in absolute terms. In terms of latency, as given in Figure \ref{fig:latency}, the FPGA platforms are generally competitive to the high performing GPU platforms. However, in terms of average power and energy use, Figures~\ref{fig:power} and \ref{fig:energy}, data centre FPGAs demonstrate their advantage. 

The utility of data centre FPGA is clearly illustrated when considering the computational efficiency of all the platforms, as given in Figure~\ref{fig:performance}, where the larger FPGAs provide many more operations per unit of energy.

A further point to consider is that Nallatech P385-D5 platform could accommodate a further 3 similar boards, where as the Max4 host system could accommodate a further 7. Doing so would significantly improve the power efficiency of both, as the energy cost of the host system would be further amortised across the FPGAs. By contrast, the GPU host systems could at most accommodate only one more GPU.

\section{Conclusion}

In this brief note, I have described a method for using data centre FPGAs. I have illustrated this method by applying it to a case study from computational finance, and evaluating it upon four data centre FPGA platforms.

As a result of this work, I conclude the following: Firstly, the potential for significantly improved power efficiency demonstrated by the data centre FPGAs in practice motivates for their existence. However, platforms should be comprised of several large FPGAs, as the host system power needs to be amortised across the computational power of the FPGAs.

Secondly, orthogonal FPGA optimisations to that what is inherent in the standard appear to yield the greatest improvements over a baseline implementation. For example, if a standard is inherently task parallel, such as OpenCL, then pipeline parallelism maximising optimisations should be employed. While counter intuitive, I believe that is due to HLS tools aggressively optimising the paradigm of the standard supported, hence requiring the programmer to explicitly flag other potential areas of optimisation.

\subsection*{Future Work}
An obvious direction future work would be to consider other workloads, such as image processing or web scale workloads, applied to this methodology. 

More ambitious future research would consider the future automation and abstraction of this method, making data centre FPGAs available to a wider audience.

\begin{acks}
Contributions from Dr David B. Thomas, Mr Shane Fleming, Prof George Constantinides and Prof Wayne Luk have been instrumental in this work. Future publications based upon this note will list them as co-authors.

Funding support has been generously provided by the Oppenheimer Memorial Trust and the South African National Research Foundation.
I would also like to thank Maxeler, Nallatech, Altera and Xilinx University Programs for supporting this work in the form of equipment and software donations. 
\end{acks}

\bibliographystyle{ACM-Reference-Format-Journals}
\bibliography{algtrad}


\end{document}